\documentclass{article}
\usepackage{slashed}

\newcommand{\bb}{\begin{eqnarray}}
\newcommand{\ee}{\end{eqnarray}}

\begin{document}

\title{Chiral gauge theories and non-abelian analogues of axions}


\author{P. Mitra\footnote{parthasarathi.mitra@saha.ac.in}\\
Saha Institute of Nuclear Physics, Calcutta 700064}

\date{}

\maketitle

\begin{abstract}
The axion particle may or may not exist, but the axion field
can be used, as shown here, in an explicitly local formulation of a 
chiral U(1) gauge theory with both classical and quantum gauge invariance.
Nonabelian analogues of axion fields, which have recently been introduced, 
can be used, together with their special symmetries, 
in a similar construction of nonabelian chiral gauge theories. 
As in known cases, the gauge symmetry is
broken and the gauge boson acquires a mass by swallowing the axions
which are therefore not physical in this construction.
\end{abstract}






\section{Introduction}
It has been known for over half a century that the chiral symmetry which holds
in classical Dirac theory with massless fermions interacting with gauge
fields is broken upon quantization \cite{abj}. The transformation 
\bb
\psi\rightarrow e^{i\alpha\gamma_5}\psi,\quad
\bar\psi\rightarrow \bar\psi e^{i\alpha\gamma_5},
\ee
is a symmetry of the kinetic term $\bar\psi[i\slashed{\partial}]\psi$
and also of the interaction term $\bar\psi[i\slashed{A}]\psi$ with the gauge
field $A_\mu$. But the axial current $\bar\psi\gamma_\mu\gamma_5\psi$, which
appears to be conserved from the equations of motion is found to violate this
conservation when the fermion triangle diagram is carefully regularized and
evaluated. This effect is the chiral anomaly. 

In the early literature,
only the one-loop diagram was considered, but subsequently it became clear
that higher order loops do not cause any further damage. Instead of considering
individual diagrams it is more natural to consider the action and regularize it
by one of the many available methods.
Calculations are done with the
temporary regularized action, after which the realistic limit is taken.
The divergence of the axial current
is proportional to $F^{\mu\nu}\tilde F_{\mu\nu}$, $F$ being the field strength.

While global symmetries can usually be made local with the addition of gauge
fields which have to be suitably transformed, this cannot be done in the
case of chiral symmetries. This is because of the same anomaly. Gauge field
equations of motion require the current to be conserved, but the anomaly
means that the relevant current is not conserved. Therefore special efforts
have to be made to define a gauge theory with a local chiral symmetry.
That anomalous gauge theories can be made sense of in two dimensions was shown
many years ago \cite{jr}. Studies of the chiral Schwinger model showed
that the photon of the two-dimensional model acquires a mass and the gauge
symmetry is broken. For more realistic models, it is usual to follow the
suggestion of \cite{fs} to reformulate an anomalous model in a gauge
invariant way by adding a group variable as an extra field.
The modified theory is not classically gauge invariant but becomes so
on quantization because of the anomaly!
We show here how axions can be used to reformulate chiral $U(1)$ gauge theories
in four dimensions in an alternative way which is manifestly
gauge invariant both before and after quantization and
then proceed to generalize to nonabelian theories.

\section{Peccei-Quinn symmetry and the axion}

Chiral symmetry is classically broken by the mass term $m\bar\psi\psi$
in addition to the quantum anomaly mentioned above. An
artificial chiral symmetry for massive fermions uses a new field
${\varphi}$ to absorb the chiral transformation. The mass term is replaced by
\cite{pq}
\bb\bar\psi m e^{i{\varphi}\gamma_5}\psi,\ee
which is invariant if the field $\varphi$ transforms under
\bb
\psi\rightarrow e^{i\alpha\gamma_5}\psi,\quad
\bar\psi\rightarrow \bar\psi e^{i\alpha\gamma_5},
\quad\varphi\rightarrow{\varphi}-2\alpha.
\label{0}\ee
This transformation leaves the action invariant provided the new field $\varphi$
is massless. This is the Peccei-Quinn symmetry. The particle \cite{ax}
corresponding to the new field $\varphi$ \cite{pq} is called the axion.
This construction was thought to solve the so-called strong CP problem. 
The particle called the axion, which was duly searched,
has not been seen in any experiment \cite{ex}. 
An alternative solution of the strong CP problem avoids the axion \cite{bcm}.
But axions are now being looked upon as examples of the hiding dark
matter and the search goes on. 

On a different front,
it has been noted recently that the Peccei-Quinn symmetry is not anomalous 
\cite{pm}, though it is spontaneously broken.
The basic technique is to introduce a regularization, say Pauli-Villars. With a
careful construction of the regulator mass term to include a coupling to the
axion, one can preserve the Peccei-Quinn symmetry.
The Lagrangian density
\bb
{\cal L}_{reg}=\bar\psi[i \slashed{D} -me^{i\varphi\gamma_5}]\psi
+\bar\chi[i \slashed{D} -M]\chi
\label{1}\ee
including a Pauli-Villars regulator field $\chi$
is invariant under (\ref{0}) but 
not under the chiral transformation of the regulator, namely
\bb
\chi\rightarrow e^{i\alpha\gamma_5/2}\chi,\quad
\bar\chi\rightarrow \bar\chi e^{i\alpha\gamma_5/2}.
\ee
That is why it was believed that the
Peccei-Quinn symmetry does not survive quantization.
However, the
regulator field must behave in the same way as the physical field. Thus the
field $\chi$ should be coupled to the gauge field $A_\mu$
and also the axion:
\bb
{\cal L}_{reg,corr}=\bar\psi[i \slashed{D} -me^{i\varphi\gamma_5}]\psi
+\bar\chi[i \slashed{D} -Me^{i\varphi\gamma_5}]\chi.
\label{2}\ee
With this correction, the action, including the kinetic term of the axion,
is invariant under the full Peccei-Quinn transformation
\bb
\psi\rightarrow e^{i\alpha\gamma_5}\psi,&\quad&
\bar\psi\rightarrow \bar\psi e^{i\alpha\gamma_5},\nonumber\\
\chi\rightarrow e^{i\alpha\gamma_5}\chi,&\quad&
\bar\chi\rightarrow \bar\chi e^{i\alpha\gamma_5},\nonumber\\
\varphi&\rightarrow&{\varphi}-2\alpha.
\ee
Thus this Pauli-Villars regularization respects the Peccei-Quinn
symmetry, which accordingly is not anomalous but survives quantization
\cite{pm}.
Note that the action is {\em not} invariant under just the chiral 
transformations of the fermion and regulator fields: the axion field also 
has to transform. That is why there is a chiral anomaly but no Peccei-Quinn
anomaly.

\section{Chiral $U(1)$ theory}

We note now that the global 
symmetry can be made local by having an appropriate gauge field $B_\mu$
for the purpose:
\bb
{\cal L}_B=\bar\psi[i \slashed{\partial}+\slashed{B}\gamma_5 -me^{i\varphi\gamma_5}]\psi
+\frac12 F^2(\partial_\mu\varphi+2B_\mu)(\partial^\mu\varphi+2B^\mu),
\ee
where
\bb
B_\mu\rightarrow B_\mu+\partial_\mu\alpha
\ee
under a local Peccei-Quinn transformation (\ref{0}). 
Note that it has to couple to both $\psi$ and $\varphi$.
Here $F$ is a constant of mass dimension such that the axion kinetic term
is $\frac12 F^2\partial_\mu\varphi\partial^\mu\varphi$ and there is an
additional kinetic term involving the gauge field strength $G_{\mu\nu}$. 
The argument for the absence of an anomaly in the Peccei-Quinn symmetry
continues to hold because the regulator mass term can again be made invariant.

A mass term is generated
for the gauge field $B$ because the symmetry is spontaneously broken.
Spontaneous breaking is clear because the transformation of $\varphi$
by an additive term cannot be consistent with any vacuum expectation value. 
The axion is the Goldstone field which vanishes in
the unitary gauge and may be considered to be
swallowed by the gauge field $B$ which becomes massive,
as may be seen by considering a local Peccei-Quinn transformation with
$\alpha=-\frac12\varphi$.
This is in fact the usual Higgs phenomenon.

Note that this is a construction of a chiral $U(1)$ gauge theory using
an axion. The presence of the axion may raise eyebrows, 
but this is similar to the reformulation used earlier \cite{fs,tsu,kra,ms,rr}.
That way of going about a chiral gauge theory
with an anomaly is to make it non-anomalous by adding a Wess-Zumino
term \cite{fs,tsu} which includes
a gauge group element as a new field. 
Instead of adding a non-invariant Wess-Zumino term to nullify the anomaly
in the quantum theory, the present approach includes the axion to
increase the classical symmetry, but this Peccei-Quinn symmetry turns out
to be free from anomaly, leading also to quantum invariance. 
The axion does not occur in the physical spectrum, just like the gauge group
element of the Wess-Zumino term: both become trivial in the unitary gauge.
In the Wess-Zumino term approach, there need not be a
kinetic term for the gauge group field, so that a gauge boson mass need not
arise when one goes to the unitary gauge. In the axion construction the mass
necessarily arises and this is as in the chiral Schwinger
model \cite{jr,grr}.
The mass of the gauge boson is not determinate in the sense that $F$ is
not fixed. In the original studies on the chiral Schwinger model too it was
undetermined, because it was related to a regularization ambiguity of the
fermion determinant. In the present approach, no determinant is calculated
and the discussion holds even in four dimensions. The photon acquires a
mass.

Note that it is also possible to make an anomalous gauge theory non-anomalous
without introducing new fields \cite{pmrr}.
There have been many studies on different aspects of chiral gauge theories.
Some recent work can be found, for example, in \cite{bra,gho}.

\section{Nonabelian chiral symmetry and nonabelian axions}
Next we consider the nonabelian generalization of this construction. First
we need a generalized axion. This is not related to its alleged connection
with the strong CP problem, but it could be relevant to dark matter.
The above axion couples to the flavour-singlet meson. As it has not been
detected so far, it is reasonable to think of non-singlet particles which could
be present in the form of dark matter \cite{nax}. 
Even if such {\it particles} do not exist, 
we may consider generalizing axion {\it fields}
for the sake of chiral gauge theory and see what consequences arise.
The abovementioned chiral symmetry is under a transformation of the fermion in
spinor space. If the fermion is an $SU(N)$ multiplet, there exist nonabelian 
chiral symmetries. The kinetic piece
\bb
\bar\psi i\slashed{\partial}\psi=
\bar\psi_L i\slashed{\partial}\psi_L +\bar\psi_R i\slashed{\partial}\psi_R
\ee
is invariant under the chiral transformations
\bb
\psi_L\rightarrow U_L\psi_L,\quad \psi_R\rightarrow U_R\psi_R,
\label{3}\ee
where $U_L,U_R$ are spacetime independent $SU(N)$ matrices acting on the
two chiral projections of $\psi$.
The gauge interactions will also be invariant under these provided
the matrix $A_\mu$ commutes with $U_L,U_R$:
if the $SU(N)$ is a flavour group, the colour $SU(3)$ could be the gauge group.

The usual mass term $m(\bar\psi_L \psi_R+\bar\psi_R \psi_L)$ is not invariant
under (\ref{3}) unless $U_L=U_R$, in which case of course the transformation is 
not really a chiral transformation.
The analogue of the Peccei-Quinn term $\bar\psi\exp(i\varphi\gamma_5)\psi$ is 
$m(\bar\psi_L W\psi_R+\bar\psi_R W^\dagger\psi_L)$,
with $W$ an $SU(N)$ matrix field.
It is analogous to the exponentiated pion field used in chiral models. But
this is different from chiral models because these are not related to quarks
and pions but are new fields.
Considering that the original $U(1)$ 
axion has not yet been detected, one may have doubts about the existence
of such new objects, but they could be looked for
in the search for dark matter. Whether they exist or not,
the idea of these fields will be theoretically useful for 
nonabelian chiral gauge theories as we shall see.

The above term can be made invariant under (\ref{3}) by transforming $W$ as \cite{nax}
\bb
W\rightarrow U_LWU^\dagger_R.
\label{4}\ee
As an $SU(N)$ matrix it involves $N^2-1$ parameters which become fields.
The kinetic term for this matrix field has to be of the form
$\frac12 F^2Tr[\partial_\mu W \partial^\mu W^\dagger]$
familiar from chiral models. This is invariant under (\ref{4}).
Here $F$ is a constant of mass dimension 1.
Thus the full action is invariant under the generalized Peccei-Quinn symmetry
\bb
\psi_L\rightarrow U_L\psi_L,\quad \psi_R\rightarrow U_R\psi_R,\quad
W\rightarrow U_LWU^\dagger_R.
\label{5}\ee

We may ask whether this generalization of nonabelian chiral symmetry
survives quantization. 
The na\"{i}ve way of regularization would be to introduce a Pauli-Villars
$SU(N)$ multiplet with the kinetic term $\bar\chi i\slashed{D}\chi$ and a mass
term $M(\bar\chi_L \chi_R+\bar\chi_R \chi_L)$. This second term
will not be invariant under chiral transformations of $\chi$.
But Pauli-Villars requires a coupling of $\chi$ to $W$:
$M(\bar\chi_L W\chi_R+\bar\chi_R W^\dagger\chi_L)$. Adding this interaction with
the $W$ field ensures invariance of the regularized action.
For the full Pauli-Villars regularization \cite{fad} one has
\bb
{\cal L}_{reg}=\bar\psi[i \slashed{D} -m(WP_R+W^\dagger P_L)]\psi
+\sum_j\sum_{k=1}^{|c_j|}\bar\chi_{jk}[i \slashed{D} -M_j(WP_R+W^\dagger P_L)]\chi_{jk},
\ee
where $P_L,P_R$ are the projection operators for left and right chirality
respectively and
$c_j$ are integers whose signs are related to the statistics assigned
to the spinor fields $\chi_{jk}$. They are positive for Fermi statistics and
negative for Bose statistics. They have to satisfy some conditions to
ensure regularization of the divergences:
\bb
1+\sum_jc_j=0,\quad m^2+\sum_jc_jM_j^2=0.
\ee
This regularized action is invariant under the transformations 
\bb
\psi_L\rightarrow U_L\psi_L, \psi_R\rightarrow U_R\psi_R,
\chi_L\rightarrow U_L\chi_L, \chi_R\rightarrow U_R\chi_R,
W\rightarrow U_LWU^\dagger_R.
\ee
Hence like the abelian Peccei-Quinn symmetry \cite{pm}, this nonabelian 
classical symmetry too survives quantization and is not anomalous.

Instead of a Pauli-Villars regularization, one may use other regularizations.
Some are discussed in \cite{nax}, and a lattice regularization \cite{wil}
is briefly discussed in the Appendix.

\section{Gauging the nonabelian Peccei-Quinn symmetry}

We shall now go on to our gauge theory. As before,
gauge fields can be used to extend the global Peccei-Quinn-like symmetries to
local ones. For example, for the left handed symmetry under $U_L$, one needs
an $SU(N)$ gauge field matrix $B_\mu$:
\bb
{\cal L}_B&=&\bar\psi[i \slashed{\partial}+\slashed{B}P_L -m(WP_R+W^\dagger P_L)]\psi\nonumber\\
&+&\frac12 F^2Tr[(\partial_\mu -iB_\mu)W (\partial^\mu W^\dagger+iW^\dagger B^\mu)]-\frac14 Tr[G_{\mu\nu}G^{\mu\nu}]
\ee
Here the transformation of the new gauge field under (\ref{5}) is given by
\bb
(\partial_\mu -iB_\mu)\rightarrow U_L(\partial_\mu -i B_\mu)U_L^{-1}.
\ee
The action is invariant under local gauge transformations $U_L$ affecting
$\psi_L, W, B$.
The right handed chiral symmetry too can be gauged if desired in a similar way.
The quantity $F$ has mass dimension one, as in the abelian case.

The Pauli-Villars regularization now reads
\bb
{\cal L}_{B,reg}&=&\bar\psi[i \slashed{\partial} +\slashed{B}P_L-m(WP_R+W^\dagger P_L)]\psi\nonumber\\
&+&\frac12 F^2Tr[(\partial_\mu -iB_\mu)W (\partial^\mu W^\dagger+iW^\dagger B^\mu)]-\frac14 Tr[G_{\mu\nu}G^{\mu\nu}]\nonumber\\
&+&\sum_j\sum_{k=1}^{|c_j|}\bar\chi_{jk}[i \slashed{\partial}+\slashed{B}P_L -M_j(WP_R+W^\dagger P_L)]\chi_{jk}.
\ee
This continues to be invariant under local $U_L$ transformations.
Thus there is no anomaly in the Peccei-Quinn type $U_L$ symmetry even now 
although there is the usual nonabelian
chiral anomaly when $W$ is not transformed.
The result of the construction is that there is formally
a gauge theory having a local chiral nonabelian
Peccei-Quinn symmetry including axions $W$.
The axions again play the r\^{o}le of gauge group valued fields in the
Wess-Zumino action \cite{fs,tsu}.

Just as in the abelian case the axion could be transformed away to zero by a
local Peccei-Quinn transformation, here too $W$ can be transformed away
to the identity by a local transformation $U_L=W^{-1}$, whereby 
\bb
W\rightarrow U_LW=1.
\ee
The Lagrangian density can then be written as
\bb
{\cal L}_{red}=\bar\psi[i \slashed{\partial}+\slashed{B}P_L -m]\psi
+\frac12 F^2Tr[B_\mu B^\mu]
-\frac14 Tr[G_{\mu\nu}G^{\mu\nu}].
\ee
Clearly, gauge invariance under
$U_L$ is lost in this reduced form and the gauge field $B$ has a mass 
whose value is undetermined because $F$ is not known.  
Global chiral symmetry is completely lost in the absence of $W$.
The nonabelian axions have in effect been swallowed by the gauge field.
These axions are thus physical only if the Peccei-Quinn symmetry is {\it not 
gauged.}

In the functional integral formulation, for each $W$ configuration,
the above $U_L$ transformation may be used to transform $W$ away to unity.
If we suppress the indices in $\chi$ and $M$ for convenience,
\bb &Z\equiv\int\int\int\int\int\int {\cal D}W{\cal D}\psi{\cal D}\bar\psi{\cal D}\chi{\cal D}\bar\chi{\cal D}B&\nonumber\\
&\exp i
\int d^4x\bigg(\bar\psi[i \slashed{\partial}+\slashed{B}P_L -m(WP_R+W^\dagger P_L)]\psi
+\bar\chi[i \slashed{\partial}+\slashed{B}P_L -M(WP_R+W^\dagger P_L)]\chi&
\nonumber\\&+\frac12 F^2Tr[(\partial_\mu -iB_\mu)W (\partial^\mu W^\dagger+iW^\dagger B^\mu)
]-\frac14 Tr[G_{\mu\nu}G^{\mu\nu}]\bigg)&
\nonumber\\
&=\int\int\int\int\int\int {\cal D}W{\cal D}\psi{\cal D}\bar\psi{\cal D}\chi{\cal D}\bar\chi{\cal D}B&\nonumber\\
&\exp i
\int d^4x\bigg(\bar\psi'[i \slashed{\partial}+\slashed{B'}P_L -m(W'P_R+W'^\dagger P_L)]\psi'
+\bar\chi'[i \slashed{\partial}+\slashed{B'}P_L -M(W'P_R+W'^\dagger P_L)]\chi'&
\nonumber\\&+\frac12 F^2Tr[(\partial_\mu -iB'_\mu)W' (\partial^\mu W'^\dagger+iW'^\dagger B'^\mu)
]-\frac14 Tr[G_{\mu\nu}G^{\mu\nu}]\bigg)&\nonumber\\
&=\int\int\int\int\int\int {\cal D}W{\cal D'}\psi{\cal D'}\bar\psi{\cal D'}\chi{\cal D'}\bar\chi{\cal D'}B&\nonumber\\
&\exp i
\int d^4x\bigg(\bar\psi[i \slashed{\partial}+\slashed{B}P_L -m]\psi
+\bar\chi[i \slashed{\partial}+\slashed{B}P_L -M]\chi&
\nonumber\\&+\frac12 F^2Tr[B_\mu B^\mu
]-\frac14 Tr[G_{\mu\nu}G^{\mu\nu}]\bigg)&\nonumber\\
&=\int\int\int\int\int\int{\cal D}W{\cal D}\psi{\cal D}\bar\psi{\cal D}\chi{\cal D}\bar\chi{\cal D}B&\nonumber\\
&\exp i\int d^4x\bigg(\bar\psi[i \slashed{D}+\slashed{B}P_L -m]\psi
+\bar\chi[i \slashed{\partial}+\slashed{B}P_L -M]\chi&
\nonumber\\&+\frac12 F^2Tr[B_\mu B^\mu
]-\frac14 Tr[G_{\mu\nu}G^{\mu\nu}]\bigg).&
\ee
In the first step, the fields are transformed by $U_L$ and the transformed
versions are indicated by primes. With the choice of $U_L$ mentioned above,
$W'=1$. The primes on the other fields are dropped in the next step,
as a change of variable,
which means that the integration measures, shown with primes,
include $U_L$: ${\cal D'}\psi={\cal D}(U_L^{-1}\psi_L){\cal D}\psi_R$ etc. But
the measure for fermions together with the Pauli-Villars fields 
${\cal D}\psi{\cal D}\bar\psi{\cal D}\chi{\cal D}\bar\chi$
is invariant under the chiral $U_L$ transformation, as in the regularized 
theory any anomaly arises from the regulator mass term \cite{bcm}.
Hence in the final step the original measure is restored. The gauge
field measure is gauge invariant in the absence of gauge fixing.
This confirms that $W$ decouples, as indicated by the local symmetry.


\section{Discussion}

To sum up, we have used axions and their nonabelian analogues
which are associated with
a nonabelian generalization of the Peccei-Quinn symmetry. The 
Peccei-Quinn symmetry is not just a classical symmetry but is non-anomalous,
as shown \cite{pm, nax} by constructing a regularization involving the axions
consistent with it. 
The abelian or nonabelian chiral Peccei-Quinn symmetry
can be made local by introducing a new gauge field. This symmetry is
spontaneously broken and the corresponding gauge field becomes massive
by swallowing the axion. Thus there emerges a new gauge invariant
axion formulation of abelian or nonabelian chiral gauge
theory {\it even if the axion does not exist physically as a particle.}
The old results of the chiral Schwinger model 
hold: the gauge symmetry is broken and the gauge
boson becomes massive. Similar results follow in four dimensions and for 
nonabelian cases. The advantages over the Wess-Zumino reformulation
are that the action is explicit, manifestly local, possesses 
gauge invariance even at the classical level and necessarily yields a mass term
for the gauge bosons.

\section*{Acknowledgment}
{It is a pleasure to acknowledge the collaboration with Debashis Chatterjee
on nonabelian axions which led to the present ideas on chiral gauge theories.}

\section*{Appendix}

The lattice regularization, which may be regarded as a kind of point-splitting
regularization, also preserves Peccei-Quinn-like symmetries of the axion
actions discussed above using Pauli-Villars regularization.

The discrete Lagrangian density for a Wilson fermion takes the form
\bb {\cal L}_{lat}=
\bar\psi[\frac12(\slashed{D}^L+\slashed{D}^{L*})-aD^{L*}_\mu D^L_\mu]\psi.
\ee
Here $a$ is the lattice spacing, $D^L_\mu$ is the covariant forward difference
(in the $\mu$-direction)
divided by $a$ and $D^{L*}_\mu$ the covariant backward difference, again
divided by $a$. While the kinetic term is chirally invariant as in the
continuum, there is the double derivative term which breaks this invariance on
the lattice \cite{wil}. 
In the axion case, the axion coupling has to be introduced at two places
in the action because there are two terms which are chirally non-invariant:
\bb
{\cal L}_1=\bar\psi[\frac12(\slashed{D}^L+\slashed{D}^{L*})
-ae^{i\varphi\gamma_5/2}(D^{L*}_\mu D^L_\mu)e^{i\varphi\gamma_5/2}
-me^{i\varphi\gamma_5}]\psi.
\ee
In the $SU(N)$ case, one needs
\bb
{\cal L}_2=\bar\psi[\frac12(\slashed{D}^L+\slashed{D}^{L*})
-a(D^{L*}_\mu D^L_\mu)WP_R-aW^\dagger P_L(D^{L*}_\mu D^L_\mu)
-m(WP_R+W^\dagger P_L)]\psi.
\ee
These lattice actions clearly preserve the Peccei-Quinn
symmetry of the continuum action when $\varphi,W$ are transformed correctly.
\end{document}